\documentclass[seceq,supplement]{ptptex}

\usepackage{graphicx}
\usepackage{wrapft}

\title{
An Efficient Joint Source-Channel Decoder with Dynamical Block Priors%
}

\author{
Ido \textsc{Kanter}, Haggai \textsc{Kfir} and  Shahar \textsc{Keren} %
}

\inst{
Minerva Center and Department of Physics, Bar-Ilan University
Ramat-Gan, 52900 Israel }

\abst{
An efficient joint source-channel (s/c) decoder based on the side
information of the source and on the MN-Gallager algorithm over
Galois fields is presented. The dynamical block priors (DBP) are
derived either from a statistical mechanical approach via
calculation of the entropy for the correlated sequences, or from
the Markovian transition matrix. The Markovian joint s/c decoder
has many advantages over the statistical mechanical approach. In
particular, there is no need for the construction and the
diagonalization of a $q \times q$ matrix and for a solution to
saddle point equations in $q$ dimensions. Using parametric
estimation, an efficient joint s/c decoder with the lack of side
information is discussed. Besides the variant joint s/c decoders
presented, we also show that the available sets of
autocorrelations consist of a convex volume, and its structure can
be found using the Simplex algorithm.

}

\begin{document}

\maketitle

\section{Introduction}
Source coding is a process for removing redundant information from
the source information symbol stream. Channel coding is a procedure
for adding redundancy as protection into the information stream
which is to be transmitted. During the last decade engineers and
also physicists have designed efficient error correction techniques
such as Low-Density-Parity-Check-Codes (LDPC)\citen{Cover,Gallager}
or Turbo codes, that nearly saturate Shannon's limit.

The Shannon separation theorem states that source coding
(compression) and channel coding (error protection) can be performed
separately and sequentially, while maintaining optimality
\citen{Shannon,Cover}. However, this is true only in the case of
asymptotically long block lengths of data and point-to-point
transmission. In many practical applications, the conditions of the
Shannon's separation theorem neither hold, even approximatly. Thus,
considerable interest has developed in various schemes of joint
source-channel (s/c) coding, where compression and error correction
are combined into one mechanism.

The paper is organized as follows. In Section 2, Statistical
Mechanical (SM) methods are used to explore properties of
correlated sequences. In section 3, the space of possible sets of
autocorrelations is investigated. In sections 4 Mackay and Neal's
(MN) algorithm for error correction is briefly introduced, and the
results of section 2 are used to extend this algorithm to a joint
s/c scheme. The estimation of the threshold of the code using the
scaling argument for the convergence time is presented in section
5. Section 6 presents simulation results, and compares joint to
separation schemes. In Sections 7 and 8, the joint s/c problem is
attacked from a different point of view, a Markovian approach, and
an efficient joint scheme \emph{with the lack of side information}
is demonstrated.


\section{Joint s/c coding - Statistical Mechanical approach}

In our recent papers\citen{KR,KK} a particular scheme based on a SM
approach for the implementation of the joint s/c coding was
presented and the main steps are briefly summarized below.  The
original boolean source is first mapped to a binary source
\citen{sourlas} $\left\{ x_{i}=\pm1\right\} ~i=1,...,L$, and is
characterized by a finite set of autocorrelations bounded by the
length $k_0$
\begin{equation}
C_{k_1, ...,k_m}=\frac{1}{L}\sum_{i=1}^{L}x_{i}\prod_{j=0}^m
x_{\left(i+k_j\right)\: \mathbf{mod}\: L} \label{ck}
\end{equation}
\noindent where $k_m \le k_0$ is the highest length autocorrelation
taken and the total number of possible different autocorrelations is
$2^{k_0}$.  For $k_0=2$, for instance, there are only $4$ possible
correlations, $C_0$, $C_1$, $C_2$ and $C_{12}$, and for $k_0=3$
there are $8$ possible different correlations;
$C_0,~C_1~,C_2,~C_3,~C_{12},~C_{13},~C_{23},~C_{123}$, where we do
not assume left-right symmetry for the source.  Note that for
general $k_0$ and $m=1$, Eq. (\ref{ck}) reduces to the two-point
autocorrelation function \citen{liat}. The number of sequences
obeying these $2^{k_0}$ constraints is given by
\begin{equation}
\Omega = Tr_{\{ x_i = \pm 1 \} }\!\!\!\!
\prod_{\{k_1,k_2,...,k_m\}}\!\!\!\!\! \delta
(\sum_{i=1}^{L}x_{i}\prod_{j=0}^m x_{i+k_j} - LC_{k_1, ...,k_m})
\label{omega}
\end{equation}
\noindent where $m=0$ stands for $C_0$. Using the integral
representation of the delta functions, Eq. (\ref{omega}) can be
written as
\begin{eqnarray}
\Omega \!\!=\!\!\!\! \int\!\!\!\! \prod_{\{k_1,..,k_m\}}
\!\!\!\!\!\!\!dy_{\{k_1,..,k_m\}} \exp(\sum
\!\!-y_{k_1,..,k_m}C_{k_1,..,k_m}) Tr \exp
(\!\!\!\sum_{k_1,..,k_m} \!\!\!\!y_{k_1,...,k_m}\!\!\!\sum_{i}
\!x_{i}\prod_{j=0}^m \!x_{i+k_j} )\nonumber \label{omega1}
\end{eqnarray}
Since $k_j \le k_0$, the last term of this equation indicates that
the trace can be performed using the standard transfer matrix (of
size $2^{k_0} \times 2^{k_0}$) method\citen{baxter}. More precisely,
assume two successive blocks of $k_0$ binary variables denoted by
$(x_1,...,x_{k_0})$ and $(x_{k_0+1},...  ,x_{2k_0})$. The element
$(i,j)$ of the transfer matrix is equal to the value of the last
exponential term (on the r.h.s of the trace), where the first block
is in state $i$ (among $2^{k_0}$ possible states) and the second
block is in state $j$. The transfer matrix is a non-negative matrix
(as long as the $y_{k_1,...,k_m}$ are real numbers), and the leading
eigenvalue is positive and non-degenerate\citen{baxter}. In the
leading order one finds
\begin{eqnarray}
\Omega =\int dy_k\exp\{-L\lbrack \sum
y_{k_1,...,k_m}C_{k_1,...,k_m}  - \ln \lambda_{max}(\{
y_{k_1,...,k_m}\}) \rbrack \} \label{omega-sp}
\end{eqnarray}
\noindent where $\lambda_{max}$ is the maximal eigenvalue of the
corresponding transfer matrix.  For large $L$ and using the saddle
point method, the entropy, $H_2(\{C_{k_1,...,k_m} \})$, is given
in the leading order by
\begin{eqnarray}
H_2\left(\{C_{k_1...,k_m}\}\right) = {1 \over \ln 2} \lbrack
\frac{1}{k_0}\ln \lambda_{max}\left (\{ y_{k_1,...,k_m}\}\right)
-\sum_{k_1,...,k_m}^{k_0}y_{k_1,...,k_m} C_{k_1,...,k_m} \rbrack
\label{entropy-ck} \label{h2}
\end{eqnarray}
\noindent where $\{y_{k_1,...,k_m}\}$ are determined from the saddle
point equations of $\Omega$ \citen{KR,KK}. Assuming a Binary
Symmetric Channel (BSC) and using Shannon's lower bound, the channel
capacity of sequences with a given set of autocorrelations bounded
by a distance $k_0$ is given by
\begin{equation}
C=\frac{1-H_{2}\left(f\right)}{H_{2}(\left\{C_{k_1,...,k_m}\}\right)-
H_{2}\left(P_{b}\right)} \label{capacity}
\end{equation}
\noindent where $f$ is the channel bit error rate and $p_b$ is a
bit error rate.  The saddle point solutions derived from Eq.
 (\ref{h2}) indicate that the equilibrium properties of the
one-dimensional Ising spin system ($x_i=\pm1$) with up to order
$k_0$ multi-spin interactions\citen{ido-msi}
\begin{equation}
H=-\sum_i \sum_{k=1}^{k_0} \frac{y_{k_1,...,k_m}}{\beta }
x_{i}\prod_{j=0}^m x_{i+k_j} \label{hamiltonian}
\end{equation}
\noindent obey in the leading order the autocorrelation
constraints, Eq. (\ref{ck}).  This property of the effective
Hamiltonian, Eq. (\ref{hamiltonian}), is used in simulations to
generate an ensemble of signals (source messages) with the desired
set of autocorrelations. {\it Note that in the following we choose
$\beta=1$, and hence we denote $\{y_{k_1,...,k_m}\}$ as
interactions.}

\section{ Possible sets of autocorrelations and the Simplex algorithm}

The entropy of correlated sequences can be calculated from Eq.
 (\ref{h2}).  For the simplest case of sequences obeying only $C_1$
and $C_2$ the numerical solution of the saddle point equations
indicate that the entropy is positive only in the regime
\begin{equation}
-(1+C_2)/2 \le C_1 \le (1+C_2)/2 \label{c1c2}
\end{equation}
\noindent where outside of this regime the entropy is zero. At the
boundaries, $C_1=|(1+C_2)/2|$, two phenomena are observed: (a) the
entropy falls abruptly to zero at the boundary, and (b) $y_1$ and
$-y_2$ diverge at the boundary (the one-dimensional Hamiltonian, Eq.
(\ref{hamiltonian}) consists of frustrated loops).

These limited results obtained from the numerical solutions of the
saddle point equations suffer from the following limitations: (a)
finding the boundaries of the region in the space of
$\{C_{k_1,...,k_m}\}$ with a finite entropy is very sensitive to
the numerical precision since on the boundary the $\{|y_i|\}$
diverge; (b) it is unclear whether the available space consists of
a connected regime; (c) the question of whether out of the space
with a finite entropy, there are a finite or infinite number of
sequences (for instance  $e^{\sqrt{L}}$) obeying the set of
autocorrelations cannot be answered using the saddle point method;
(d) extension of the saddle point solutions to identify the
boundaries of the finite entropy regime to many dimensions is a
very heavy numerical task.

To overcome these difficulties, we show below how the possible sets
of autocorrelations can be identified using the Simplex algorithm.

For the case of only two constraints $C_1$ and $C_2$, for instance,
following the methodology of the transfer matrix, let us concentrate
on four successive binary variables $S_i,S_{i+1},S_{i+2},S_{i+3}$,
where $S_i=\pm 1$. Since the Hamiltonian, Eq. (\ref{hamiltonian}),
obeys in this case an inversion symmetry, let us examine only the
$8$ configurations out of $16$ where $S_3=-$, $(\pm\pm-\pm)$. For
these $8$ configurations one can assign the following marginal
probabilities, $P_{\pm\pm-\pm}$, where each probability stands for
the fraction of sequences obeying $C_1$ and $C_2$ with a given state
for these four successive binary variables. In the SM language we
measure the probabilities of these four states in thermal
equilibrium of the micro-canonical ensemble obeying Eq.  (\ref{ck}).
It is clear that the Hamiltonian, Eq. (\ref{hamiltonian}), is
translationally invariant, $P(S_i,S_{i+1},S_{i+2},S_{i+3})$ is
independent on $i$ after averaging over all sequences obeying
constraints (\ref{ck}).

For these $8$ marginal probabilities one can write the following
$14$ equations:

\begin{eqnarray}
&P&_{---+}+P_{----}+P_{++-+}+P_{++--}-P_{-+-+}-P_{-+--}-P_{+--+}-P_{+---}=C_1/2\nonumber\\
&P&_{---+}+P_{----}+P_{+---}+P_{+--+}-P_{-+-+}-P_{-+--}-P_{++--}-P_{++-+}=C_1/2\nonumber\\
&P&_{-+--}+P_{----}+P_{+---}+P_{++--}-P_{-+-+}-P_{---+}-P_{+--+}-P_{++-+}=C_1/2\nonumber\\
&P&_{---+}+P_{----}+P_{-+--}+P_{-+-+}-P_{+--+}-P_{+---}-P_{++--}-P_{++-+}=C_2/2\nonumber\\
&P&_{----}+P_{-+-+}+P_{+---}+P_{++-+}-P_{---+}-P_{-+--}-P_{+--+}-P_{++--}=C_2/2\nonumber\\
&P&_{----}+P_{---+}+P_{-+--}+P_{-+-+}+P_{+---}+P_{+--+}+P_{++--}+P_{++-+}=1/2\nonumber\\
&0& \le P_{\pm\pm-\pm} \le 1 \label{simplex12}
\end{eqnarray}

\noindent For a given $C_1$, these $14$ equations can be solved for
the minimum and the maximum available $C_2$ using the Simplex
method. Running over values of $-1 \le C_1 \le 1$, we indeed recover
the result of Eq. (\ref{c1c2}). However, the {\it Simplex solution
indicates the lack of even a finite number of sequences beyond the
regime with finite entropy}. Hence, a simple geometrical calculation
obeying constraint \ref{c1c2} indicates that the fraction of the 2D
space (-1:1,-1:1) of $(C_1,C_2)$ with available sequences is $1/2$.

For the case of three constraints, $C_1, C_2$ and $C_3$, one can
similarly write  $45$ equalities and inequalities for the $32$
probabilities of $6$ successive binary variables
$P_{\pm\pm\pm\pm-\pm}$. For a given $C_1$ and $C_2$, these $45$
equations and inequalities can be solved for the minimum and the
maximum available $C_3$ using the Simplex method. The Simplex
solution indicates: (a) the solution space in the three-dimensional
box $(-1:1,-1:1,-1:1)$ for $(C_1,C_2,C_3)$ is a connected region
bounded by a few planes. This result is consistent with the solution
obtained from Eqs. (\ref{omega-sp},\ref{h2}); (b) the fraction of
the volume of the box obeying the three constants is $\sim 0.222$.
Preliminary results indicate that for $4$ ($C_i,~i=1,2,3,4$) and $5$
($C_i,~i=1,2,3,4,5$) constraints the available volume is $\sim
0.085,~0.034$, respectively.

The fraction of possible sets of autocorrelations appears  to
decrease as the number of constraints increases. However, the
question of whether the fraction of available autocorrelations
drops exponentially with the number of constraints as well as its
detailed spatial shape is the subject of our current research.

We conclude the discussion in this section with the following
general result. The available volume for the general case of $q$
constraints $\{C_{k_1,...,k_m}\}$, $k_m<\log_2(q)$, is convex. The
main idea is that one can verify that the set of equalities can be
written in a matrix representation in the following form
\begin{equation}
{\bf M} P = C \label{manfred}
\end{equation}
\noindent where ${\bf M}$ is a matrix with elements $\pm 1$; $P$
represents the marginal probabilities $P(\pm,\pm,....)$ and $C$
represents the desired correlations or a normalization constant (for
instance $C_1/2$, $C_2/2$ and $1/2$, for the case of Eq.
(\ref{simplex12})). The inequalities force the probabilities into
the range $\lbrack 0:1\rbrack$. Clearly if $P_1(\pm,\pm,...)$ and
$P_2(\pm,\pm,...)$ are two sets of probabilities obeying Eq.
 (\ref{manfred}) then
\begin{equation}
\lambda P_1+(1-\lambda)P_2 \label{convex}
\end{equation}
\noindent is also a solution of the set of the equalities ($0 \le
\lambda \le 1$).  Hence, the available volume is convex.




\section{Joint s/c decoder: Statistical Mechanical approach}
The transfer matrix method indicates that the relevant scale of the
correlated source message is $k_0$.  Hence, our encoding/decoding
procedure is based on the MN code for a finite field $q=2^{k_0}$
\citen{LDPC-GF(q)}, which is based on the construction of two sparse
matrices $A$ and $B$ of dimensionalities $(L_0/R)\!\times\! L_0$ and
$(L_0/R)\! \times\! (L_0/R)$ respectively, where $R$ is the
code-rate and the number of symbols in the source is $L_0=L/k_0$.
The matrix $B^{-1}A$ is then used for encoding the message
\begin{equation}
t = B^{-1}A x\  (\: \mathbf{mod}\:\ \ q) \label{trans}
\end{equation}
The finite field message vector $t$ is mapped to a binary vector
and then transmitted. The received message, $r$, is corrupted by
the channel bit error rate, $f$.

The decoding of symbols of $k_0$ successive bits (named in the
following as a {\it block} of bits or binary variables) is based
on the solution of the syndrome
\begin{equation}
Z= Br = Ax + Bn\   (\: \mathbf{mod}\:\ \ q) \label{decoding}
\end{equation}
\noindent where $n$ stands for the corresponding noise of $k_0$
successive bits.  The solution of the $L_0/R$ equations with
$L_0(1/R+1)$ variables is based on the standard message passing
algorithm introduced for the MN decoder over Galois fields with
$q=2^{k_0}$\citen{LDPC-GF(q)} and with the following modification.
The horizontal pass is left unchanged, {\it but a dynamical set of
probabilities assigned for each block is used in the vertical pass}.
The Dynamical Block Probabilities (DBP), $\{P_n^c\}$, are determined
following the current belief regarding the neighboring blocks and
are given by
\begin{eqnarray}
\gamma_{n}^{c}  =  S_{I}\left(c\right)\left(\sum
_{l=1}^{q}q_{L}^{l}S_{L}\left(l,c\right)\right)\left(\sum
_{r=1}^{q}q_{R}^{r}S_{R}\left(c,r\right)\right)  ~~~~ ;~~~~
P_{n}^{c} = \frac{\gamma _{n}^{c}}{\sum _{j=1}^{q}\gamma
_{n}^{j}}\label{tm-vertical-pass} \label{dbp}
\end{eqnarray}
\noindent where $l/r/c$ denotes the state of the left/right/center
($n\!-\!1\,/\,n\!+\!1\,/\,n$) block respectively and
$q_{L}^{l}/q_{R}^{r}$ are their posterior probabilities.
$S_I(c)=e^{-\beta H_I}$ stands for the Gibbs factor of the inner
energy of a block ($k_0$ successive binary variables spins),
characterized by an energy $H_I$ at a state $c$, see Eq.
 (\ref{hamiltonian}).
Similarly $S_L(l,c)$ ($S_R(c,r)$) stands for the Gibbs factor of
consecutive Left/Center (Center/Right) blocks at a state $l,c$
$(c,r)$ \citen{KK,KR}. The complexity of the calculation of the
block prior probabilities is $O(Lq^2/ \log q)$ where $L/\log q$ is
the number of blocks.  The decoder complexity per iteration of the
MN codes over a finite field $q$ can be reduced to order
$O(Lqu)$\citen{David-Mackay1}, where $u$ stands for the average
number of checks per block.  Hence the total complexity of the DBP
decoder is of the order of $O(Lqu+Lq^2/ \log q)$.


\begin{figure}[htb]
\parbox{\halftext}{
\includegraphics[width=2.0in]{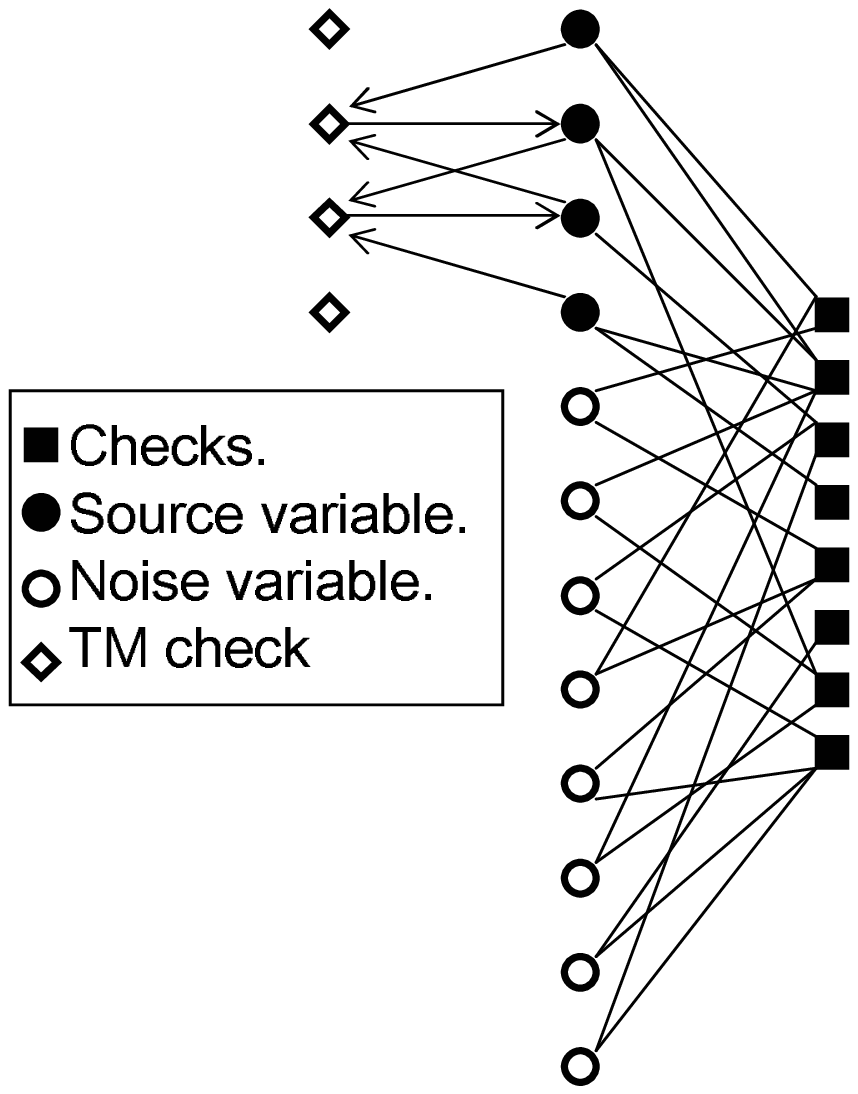}
\caption{A message passing in the joint s/c decoder is represented
by a bipartite graph with an additional
layer.}\label{message-passing}} \hfill
\parbox{\halftext}{
\includegraphics[width=2.5in]{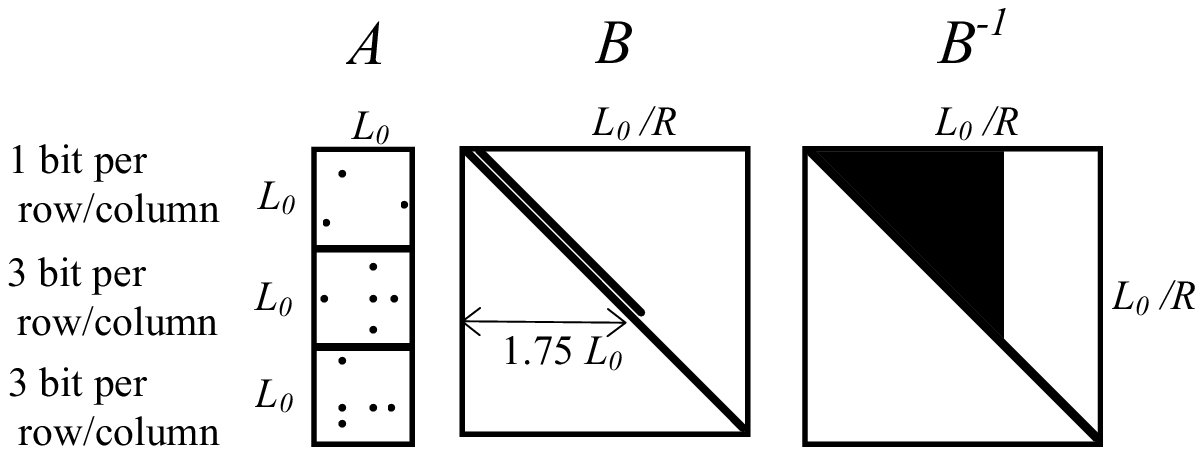}
\caption{The structure of the matrices $A$ and $B$ for the MN
decoder taken from reference\citen{KS}, for rate $1/3$. The black
dots (area) denote the non-zero elements of the matrices
$A,~B,~B^{-1}$. } \label{ks}}
\end{figure}

Another way to represent the dynamical behavior of the SM joint
s/c decoder is in the framework of message passing on a graph
(Fig. \ref{message-passing}). Typically for LDPC, the graph is
bipartite and consists of variable nodes (circles) and check nodes
(squares). For the MN algorithm, there are two types of variable
nodes: source nodes (filled circles) and noise nodes (open
circles). A message from variables to checks is a horizontal pass,
and a message from checks to variables is a vertical pass. In the
joint s/c decoder there is a {\it third} layer (diamonds): each
element in this layer sends a message (outbound arrow) to a single
source variable (namely, the dynamical block prior), and receives
two messages (inbound arrows) from the neighboring source
variables (namely their a-posteriori probabilities).


For simplification of the discussion below, in almost all of the
simulation results we concentrate on rate $1/3$ and the construction
of the matrices $A$ and $B$ follow reference \citen{KS} which is
sketched in Fig. \ref{ks}. The advantage of this construction is
that the matrices $A$ and $B$ are very sparse, but the threshold of
the code for large blocks is only $1-3\%$ of the channel
capacity\citen{KS,KS-Gaussian}. Furthermore, since $B$ has a
systematic structure, the complexity of the encoder scales linearly
with $L$ although $B^{-1}$ is dense\citen{saad,saad1}. Of course,
codes with higher thresholds exist, hence the performance of the
joint s/c algorithm reported below should be interpreted as a lower
bound. (Results for a limited example with rate greater than one,
$R>1$, are briefly discussed in reference \citen{r89})

We conclude this section with the comment that the possibility of
the SM joint s/c algorithm in the framework of the MN-Gallager
decoder to the Gallager decoder\citen{Gallager} is in question. In
the Gallager decoder we first solve $L_0(1/R-1)$ equations for the
noise variables, and only in the final step is the message
recovered. Since the noise is not spatially correlated, we do not
see a simple way to incorporate in the Gallager case the side
information about the spatial correlations among the message
variables. The equivalence between these two (MN-Gallager and
Gallager) similar decoders therefor also in doubt.

\begin{figure}[htb]
\parbox{\halftext}{
\includegraphics[height=2.5in, angle=270]{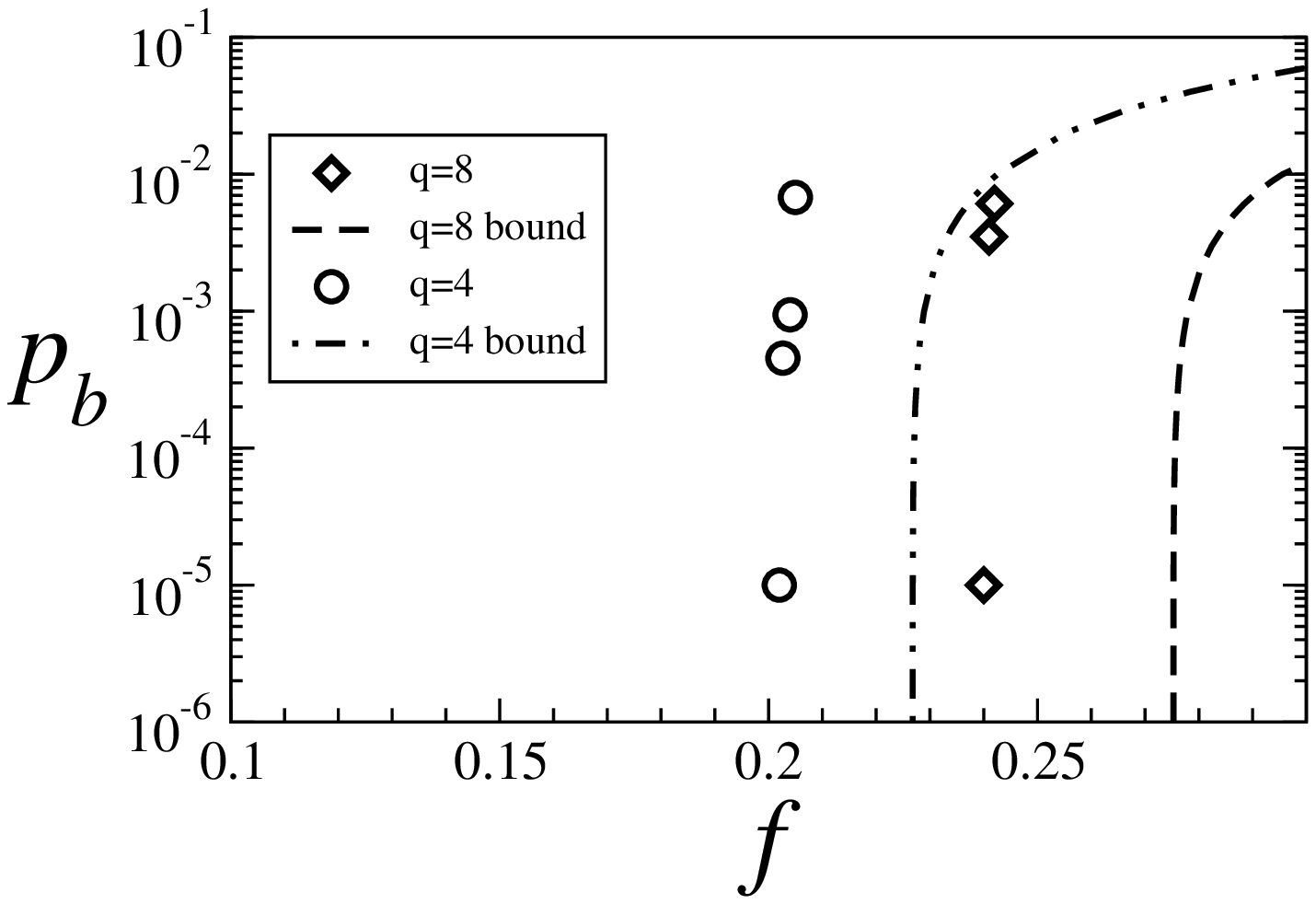}
\caption{Simulation results for rate $R=1/3$, $L=10,000$, $q=4$
and $8$. Each point was averaged over at least $1,000$ source
messages with the desired set of autocorrelations. (refer to text
for description)} \label{msi_pb} }
 \hfill
\parbox{\halftext}{
\includegraphics[width=2.0in,height=2.0in,angle=270]{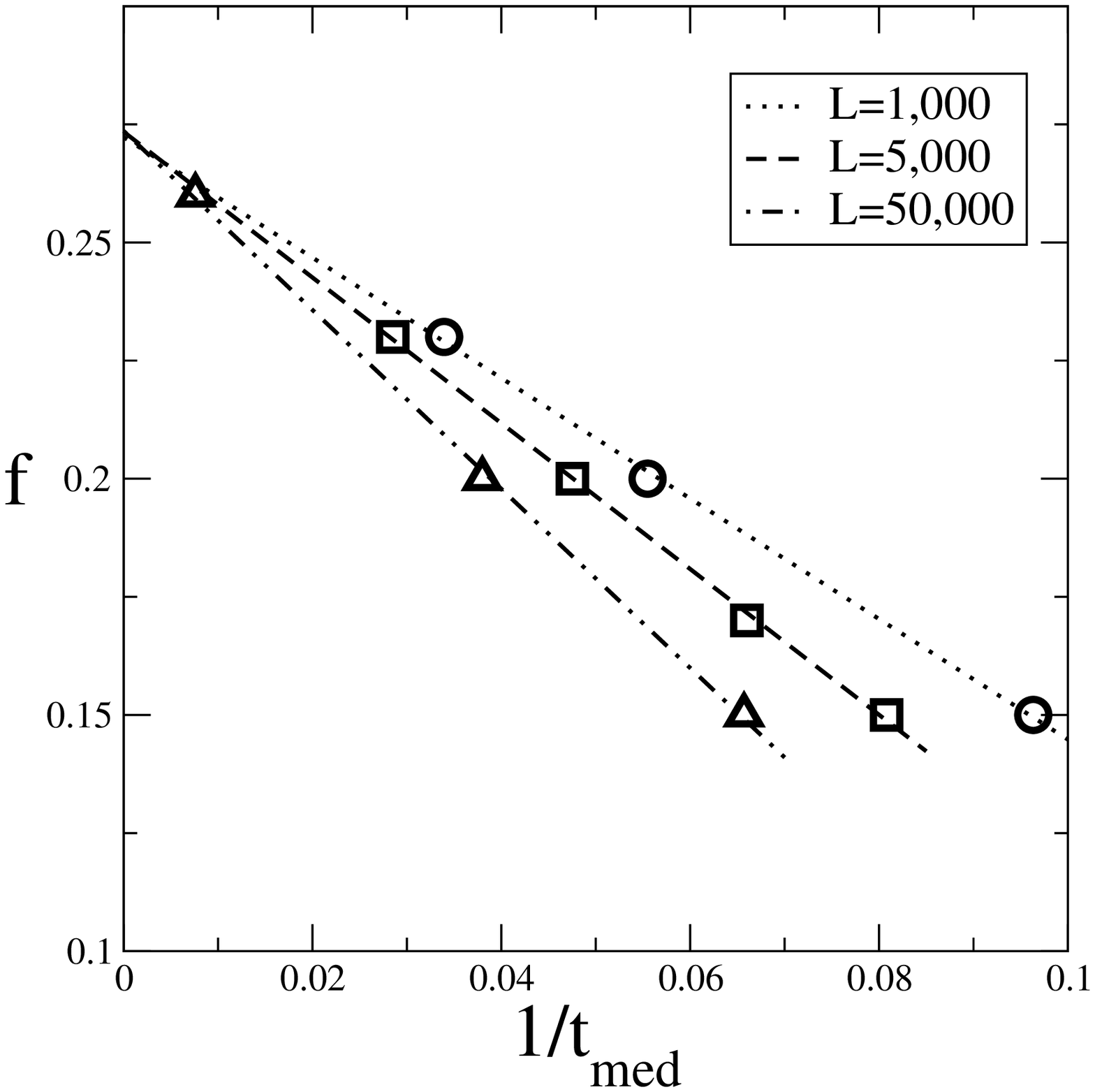}
\caption{The flip rate $f$ as a function of $1/t_{med}$ for
$GF(4)$ with $C_1=C_2=0.8$ and $L=1,000,~5,000~,50,000$.  The
lines are a result of a linear fit. The threshold, $f_{\infty}
\sim 0.272$, extrapolated from the scaling behavior Eq.
(\ref{scaling}), is independent of $L$.  } \label{different_N} }
\end{figure}

For illustration, in Fig. \ref{msi_pb} we present results for rate
$R=1/3$, $L=10,000$, $q=4$ and $8$ where the decoding is based on
the dynamical block posterior probabilities, Eq. (\ref{dbp}), and
with the following parameters. For $q=4$ (open circles)
$C_1=0.55,~C_2=0.5,~C_{12}=0.4$
($y_1=0.275,~y_2=0.291,~y_{12}=0.149$) and $H_2=0.683$. Shannon's
lower bound, Eq. (\ref{capacity}), is denoted by the double dotted
line, where for $p_b=0$ the channel noise level is $f_c=0.227$. For
$q=8$ (open diamonds) $C_1=0.77,~C_2=0.69,~C_3=0.56,~C_{123}=0.7$
($y_1=0.349,~y_2=0.36,~y_3=-0.211,~y_{123}=0.443$) and $H_2=0.453$.
Shannon's lower bound is denoted by the dashed line, where for
$p_b=0$ the channel noise level is $ f_c=0.275$.  Each point was
averaged over at least $1,000$ messages. These results for both
$q=4$ and $8$ indicate that the threshold of the presented decoder
with $L=10,000$ is $\sim 15\%-20\%$ below the channel capacity for
infinite source messages. It is worth mentioning, that without using
the dynamical block priors, Eq. (\ref{dbp}), the decoder fails to
decode at $f_c=\sim 0.13-0.14$.

\section{The threshold of the code}

The threshold $f_c$ for $L \rightarrow  \infty $ is estimated from
the scaling argument of the convergence time, which was previously
observed for $q=2$\citen{KS,KS-Gaussian}.  The median convergence
time, measured in iterations of the MN algorithm, is assumed to
diverge as the level of noise approaches $f_c$ from below. More
precisely, we found that the scaling for the divergence of $t_{med}$
is {\it independent of $q$} and is consistent with:
\begin{equation}
t_{med} = {A \over f_c-f} \label{scaling}
\end{equation}
\noindent where for a given set of autocorrelations and $q$, $A$ is
a constant. Moreover, for a given set of autocorrelations and a
finite field $q$, the extrapolated threshold $f_c$ is independent of
$L$, as demonstrated in Fig. \ref{different_N}.  This observation is
essential to determine the threshold of a code based on the above
scaling behavior. Note that the estimation of $t_{med}$ is a simple
computational task in comparison with the estimation of low bit
error probabilities for large $L$, especially close to the
threshold. We also note that the analysis is based on $t_{med}$
instead of the \emph{average} number of iterations,
$t_{av}$,\citen{KS} since we wish to prevent the dramatic effect of
a small fraction of finite samples with slow convergence or no
convergence.\citen{median} We note that preliminary results
indicates that for a given $\{C_k\}$, $f_c(q)$ appears
asymptotically to be consistent with $f_c(q) \sim f_c -const/q$.



\section{Comparison between joint and separation schemes}

\begin{figure}
\centering
\includegraphics[width=5.75in]{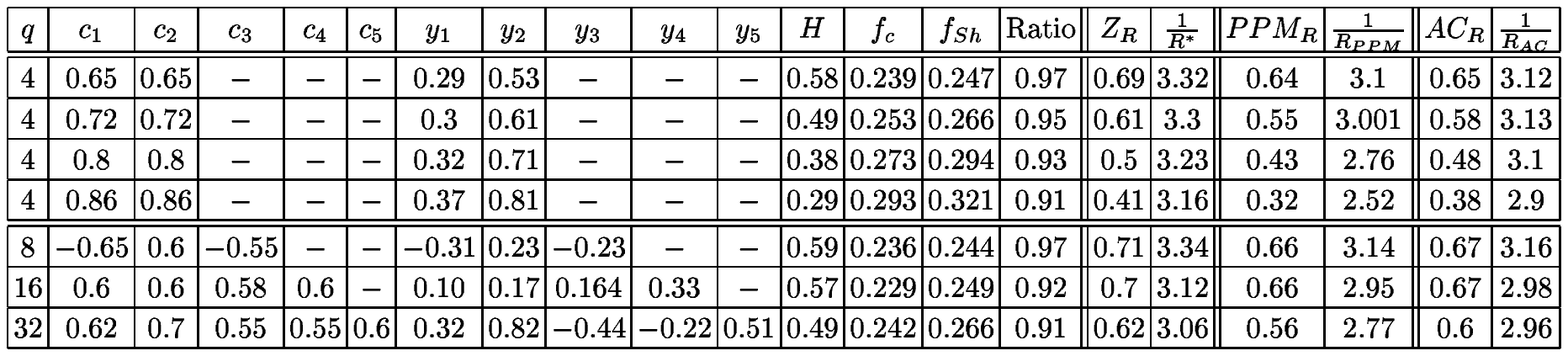}
\caption{Results for $q=4,~8,~16,~32$ and selected sets of
two-point autocorrelations $\{C_k\}$ } \label{t1}
\end{figure}

Results of simulations for $q=4,~8,~16$ and $32$ and selected sets
of autocorrelations are summarized in Table I (Fig. \ref{t1}) and
the definition of the symbols is: $\{C_k\}$ denotes the imposed
values of two-point autocorrelations as defined in eqs.  \ref{ck}
and \ref{omega}; $\{y_k\}$ are the interactions strengths, Eq.
 (\ref{hamiltonian}); $H$ represents the entropy of sequences with
the given set of autocorrelations, Eq. (\ref{entropy-ck}); $f_c$ is
the estimated threshold of the MN decoder with the DBP derived from
the scaling behavior of $t_{med}$, Eq. (\ref{scaling}); $f_{Sh}$ is
Shannon's lower bound, Eq. (\ref{capacity}); Ratio is the efficiency
of our code, $f_c/f_{Sh}$; $Z_R$ indicates the gzip compression rate
averaged over files of the sizes $10^5-10^6$ bits with the desired
set of autocorrelations. We assume that the compression rate with
$L=10^6$ achieves its asymptotic ratio, as was indeed confirmed in
the compression of files with different $L$; $1/R^{\star}$ indicates
the ideal (minimal) ratio between the transmitted message and the
source signal after implementing the following two steps:
compression of the file using gzip and then using an {\it ideal
optimal encoder/decoder}, for a given BSC with $f_c$.  A number
greater than (less than) $3$ in this column indicates that the MN
joint s/c decoder is more efficient (less efficient) in comparison
to the channel separation method using the standard gzip
compression.  The last four columns of Table I (Fig. \ref{t1}) are
devoted to the comparison of the presented joint s/c decoder with
advanced compression methods. $PPM_R$ and $AC_R$ represent the
compression rate of files of the size $10^5-10^6$ bits with the
desired autocorrelations using the Prediction by Partial
Match\citen{PPM} and for the Arithmetic Coder\citen{AC},
respectively. Similarly to the gzip case, $1/R_{PPM}$ and $1/R_{AC}$
denote the optimal (minimal) rate required for the separation
process (first a compression and then an ideal optimal
encoder/decoder) assuming a BSC with $f_c$.

Table I indicates the following main results: (a) a degradation in
the performance is observed as the correlations are enhanced. (b)
The efficiency of our joint s/c coding technique is superior to the
alternative standard gzip compression in the separation technique.
The gain disappears as the entropy decreases. These results are
farther discussed in \citen{KK} .
%

The DBP decoder based on the SM approach suffers one major
disadvantage: computing the ${y_k}$ interactions, which involves
finding the eigenvalues of a $q\times q$ matrix ($O(q^3)$), and
solving the saddle-point equations, \ref{h2}, which is also a
heavy numerical task, even for $k_0>4$.

\section{Markovian joint s/c decoder}

In order to overcome the abovementioned  drawbacks of the SM
approach, we now treat the source sequence from a different point
of view, by assuming that these sequences were generated by a
Markov process. Hence, the sequence can be described by an
alphabet $GF(q)$, a transition matrix $T_{kj}=P(x_i=j|x_{i-1}=k)$,
and the stationary solutions of the process, $P(j)$. The key point
is that for large messages, $T$ and $P(j)$ can be (approximately)
measured by the sender (for each chunk of data) with $O(L_0)$
operations. $T$ and $P(j)$ may then be transmitted reliably to the
decoder and used as side information in the decoding process
(similar to the transmission of the ${y_1,...y_{k_0}}$
interactions in the SM approach).

Consider three successive symbols $x_{i-1},x_i,x_{i+1}$ in such a
sequence. The probability of the triplet $a,b,c$ is given by:

\begin{eqnarray}\label{Pabc}
  P(a,b,c)=P(a,b)\cdot P(c|a,b)=P(a,b)\cdot P(c|b)
=\frac{P(a,b)P(b,c)}{P(b)}
\end{eqnarray}

\noindent where use has been made of the Bayes Rule:
$P(x,y)=P(x)\cdot P(y|x)$, and fact that the process is
memoryless. Now, given the a-posteriori probabilities for the
first and last symbols in the triplet: $q_{i-1}^a=Pr(x_{i-1}=a)$
and $q_{i+1}^c=Pr(x_{i+1}=c)$, one can calculate a dynamical block
prior (corresponding to the prior in (\ref{dbp}) for the
probability that $x_i=b$:

\begin{eqnarray}\label{Prior(b)}
  Pr(x_i=b)=\frac{1}{Z}\cdot \sum_{a,c=1}^q{P(a,b,c)\cdot q(a)}\cdot q(c)=\nonumber \\
  =\frac{1}{Z}P(b)^{-1} \left(\sum_{a=1}^q{P(a,b) q_{i-1}^a}\right)\cdot
 \left(\sum_{c=1}^q{P(b,c) q_{i+1}^c}\right),
\end{eqnarray}

\noindent where $Z$ is a normalization constant such that:
$\sum_{b=1}^q{Pr(x_i=b)}=1$.

The extension of the MN algorithm to the joint source-channel case
consists of the following steps:\begin{enumerate}
    \item A binary sequence of $L_0\cdot log_2(q)$ bits is converted to $L_0$ $GF(q)$ symbols.
    \item The encoder measures $T$ and $P(j)$ for all
    the $q$ symbols over the source, and transmits reliably this side
    information to the decoder.
    \item The source is encoded according to
    (\ref{trans}), then reconverted to binary representation
    and transmitted.
    \item The decoder maps the received signal back to GF($q$), and
    performs the regular decoding
    (\ref{decoding}), but after every iteration of message passing, the prior for each source symbol is
    recalculated according to (\ref{Prior(b)}).
\end{enumerate}
The complexity of calculating the $q$ priors for a single symbol
according to the posteriors of its neighbors is reduced from $q^3$
to $q^2$ by Eq. (\ref{Prior(b)}), hence the decoder's complexity
remains linear, with total complexity of $O(L_0qu+L_0q^2)$ per
iteration.

\begin{wrapfigure}{r}{6.6cm} 
\includegraphics[width=2.5in, angle=270]{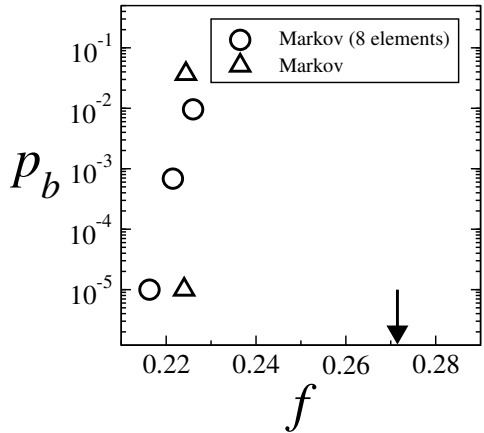}
\caption{ The bit error rate, $p_b$ versus the channel bit error
rate $f$ for $L=10,000$, $R=1/3$, $q=8$ with $C_1=C_2=C_3=0.7$.
Decoding following the Markovian process, Eq. (\ref{Prior(b)})
(open triangle), decoding following the Markovian process where
only $8$ dominated elements of the transition matrix, $T$, are
taken as a side information (open circle). Shannon's lower bound,
$f_c=0.271$ ($H=0.47$), is denoted by an arrow.} \label{q88}
\end{wrapfigure}

However, there is still a need for the transmission of the side
information ($T$, $P(j)$). Hence the size of the header is of the
order of $O(q^2)$.  For $L \rightarrow \infty$ or more precisely
for $L \gg q^2$ the overhead of the transmitted side information
is negligible; however, for a finite $L \le q^2$ it may cancel the
benefits of the Markovian joint s/c decoder.

The header size can be reduced using the following observation:
For sequences with enhanced autocorrelations, the structure of $T$
was observed to be dominated by a small number of large elements.
One can transmit only these dominated elements, and the remaining
elements of each row of $T$ will be filled evenly to maintain:
$\sum_{j=1}^q{T_{ij}}   =1$.

Fig. \ref{q88} represents simulation results for the Markovian
scheme.  Triangles mark decoding using full $T$, while circles
mark decoding using only the 8 (out of 64) most dominant elements
of $T$. The performance seems to be only slightly affected by this
approximation, which dramatically reduces the required transmitted
side information.

%

An interesting open question is the effect of use of the
sparseness of $T$ on  the complexity and performance of the
decoder.

\section{Efficient Joint s/c decoder with the lack of side information}

The discussion in the previous sections indicates that the
performance of the presented joint s/c coding is not too far from
Shannon's lower bound and, most probably, using an optimized code
(a better construction for the matrices $A$ and $B$ of the MN
code), the channel capacity can be nearly saturated. However, for
a finite block length. the main drawback of our algorithm is the
overhead of the header which must be encoded and transmitted
reliably. One has to remember that the size of the header scales
with $q^2$ where the precision of each element is of the order
$O(\log L)$. This overhead is especially intolerable in the limit
where $q^2 \log / L \sim O(1)$
Note that this is indeed the situation even for very large
messages, $L=10^6$, and the largest taken autocorrelation length
is only $\log_2q=8$.
%

In this section we explain how the Markovian joint s/c can be
implemented without any side information. The key points are the
special properties of KS construction \citen{KS} (Fig. \ref{ks}):
the first $L_0$ rows of $A$ are characterized by one non-zero
element per row and column, where the first $L_0$ rows of $B$ are
characterized by $2$ non-zero elements. Furthermore, due to the
systematic form of $B$, each row cannot be written as a linear
combination of the other rows. Hence, the first $L_0$ bits of the
syndrome, Eq. (\ref{decoding}), are equal (up to a simple
permutation) to the source, with an effective flip rate, $f_{eff}$.
For $GF(2)$ for instance, $Z_j=x_i+n_j+n_{j+1}$ ($i$ marks the
position of the nonzero element in the $j^{th}$ row of $A$), and
$f_{eff}=2f(1-f)$. The first $L_0$ symbols of $Z$ are therefore a
result of a Hidden Markov Model (HMM).  The underlaying transition
matrix, $T$, generating the source sequence, can be estimated by
means of the EM algorithm \citen{EM}. This is a standard tool for
solving such \emph{Parametric Estimation} problems, which has linear
complexity. Having $T$ (approximately) revealed, the Dynamical Block
Priors can be used as described in \ref{Prior(b)}.

For the general construction of the MN algorithm, one adds/subtracts
rows of the concatenated matrix $[A,B]$ and the corresponding
symbols in $z$ (see Eq. (\ref{decoding})), such that the following
situation is finally reached: The first $L_0$ rows of $A$ are the
identity matrix, regardless of the construction of the first $L_0$
rows of $B$. From the knowledge of the noise level $f$ and the
structure of $i^{th}$ row of $B$ one can now calculate the effective
noise level, $f_{i,eff}$, of the $i^{th}$ received source symbol.
Since all $\{f_{i,eff}\}$ are functions of a unique noise level $f$,
one can again estimate the parameters of the Markovian process using
some variants of the EM algorithm. Note, that in the general case
the first $L_0$ rows of $B$ contain loops, hence the $\{f_{i,eff}
\}$, are correlated. However, these correlations are assumed to be
small as the typical loop size is of the order of $O(\log
(L))$\citen{erdos}.

\section{Concluding remarks}

The only remaining major drawback of the presented Markovian joint
s/c coding is that the complexity of the decoder per message
passing, scales as $O(Lq^2/\log_2(q))$, this may considerably slow
down the decoder even for moderate alphabet size. Note however,
that for large $q$, such that $q^2\geq L$, and low entropy
sequences, the transition matrix, $T$, is expected to be very
sparse (consider $q=1024$ vs. block size of $L=100,000$). Taking
advantage of the sparseness of $T$, the complexity of the decoder
can be further reduced.

The one-dimensional Markovian joint s/c decoder can be easily
extended to coding of a two-dimensional array of symbols or even to
an array of symbols in higher dimensions. The complexity of the DBP
decoder scales as $L_0^dq^{2d+1}$, where $L_0^d$ is the number of
blocks in the array, and $d$ denotes the dimension\citen{haggai1}.
Using Markovian and Bayesian assumptions, the complexity can be
reduced to $O(L_0^dq^2)$.

\section*{Acknowledgements}
I.K. thanks David Forney, Wolfgang Kinzel, Manfred Opper, Shlomo
Shamai, and Shun-ichi Amari for many helpful discussions and
comments. This work was partially supported by the Israeli Academy
of Science.

%


\begin{thebibliography}{99}

%
%
%
\bibitem{Cover} Cover TM, Thomas JA. {\it Elements of information
theory.}  Wiley. 1991, UK.

\bibitem{Gallager} Gallager RG,  {\em Low Density Parity Check Codes}
Research monograph series {\bf 21} (MIT press), 1963.

\bibitem{Shannon} Shannon CE, A mathematical theory of
communication, {\emph Bell System Technical J.}, {\bf 27},
379-423, 623-656, 1948.

\bibitem{KR} Kanter I and Rosemarin H, (cond-mat-0301005).

\bibitem{KK} Kanter I and Kfir H,
 {\it Europhys. Lett. Vol. 63 No. 2
pp. 310 (July 2003)}.

\bibitem{sourlas} Sourlas N,
{\it Nature, vol.339, no.6227, 29 June 1989, pp.693-5}.

\bibitem{liat} Ein-Dor L, Kanter I, Kinzel W,
  {\it Physical Review E, vol.65, no.2, Feb.
2002}.

\bibitem{baxter}Baxter RJ, Exactly Solved Models in Statistical
Mechanics, {\emph Academic Press, London}, 1982.

\bibitem{ido-msi} Kanter I,
{\it J. Phys. A, vol. 20 pp. L257 1987}.

\bibitem{LDPC-GF(q)} Davey MC and MacKay DJC,
{\it Communications Letters, vol.2, no.6, June 1998, pp.165-7}.

\bibitem{David-Mackay1} MacKay DJC and Davey MC,
{\it Gallager Codes for Short Block Length and High Rate
Applications, Codes, Systems and Graphical Models}, IMA Volumes in
Mathematics and its Applications, Springer-Verlag (2000).

\bibitem{KS} Kanter I, Saad D, {\em Phys. Rev. Let., vol.83,
no.13, 27 Sept. 1999, pp.2660-3.}

\bibitem{KS-Gaussian} Kanter I, Saad D,
{\it j. Phys. A, vol.33, no.8, 3 March 2000, pp.1675-81.}

\bibitem{saad} Kabashima Y, Saad D,
{\it Europhysics Letters, vol.45, no.1, 1 Jan. 1999, pp.97-103}.

\bibitem{saad1} Skantzos NS, van Mourik J,  Saad D and
Kabashima  Y,
J. Phys. A 36 No 43 (31 October 2003) 11131-11141.

\bibitem{r89} For rate $9/8$, for instance, the chosen construction
for the matrices $A$ and $B$ is as follows. $A$ hase sets of rows
with $1,2,3,5,9$ random non-zero elements per row. the number of
rows from each type is $(m/16,m/4,m/16,m/16,m/16)$, where $m$ is
the number of rows in $A$. The structure of $B$ is the same as
illustrated in Fig. \ref{ks}, but $1.75$ is replaced with $7/9$.
We ran simulations for this construction with $C_1=C_2=0.7$ and
the corresponding entropy is $H_2=0.513$ and $L=9,000$.  The
extrapolation of $t_{med}$ indicates that the threshold of this
code for large $L$ is $f_c \sim 0.057$. In the separation scheme
using {\it optimal compression and error correction schemes} and
with $f_c=0.057 ~(R_{i.i.d}=0.618)$, one can find that the overall
inverse rate of the communication channel is $1/R=0.513/0.618 \sim
0.83$, which is only about $6\%$ below our joint s/c inverse rate
$1/R=8/9 \sim 0.89$.  One must remember that our MN construction
can be further optimized, and the critical channel noise is
expected to be enhanced, $f_c > 0.057$.

\bibitem{median} In practice we define $t_{med}$ to be the average
convergence time of all samples with $t \le$ the median time.

\bibitem{PPM} The PPMZ software used can be downloaded from
www.cbloom.com/src/ppmz.html

\bibitem{AC} The AC software used can be downloaded from
www.cs.mu.oz.au/~alistair/arith\_coder

\bibitem{EM} McLachlan GJ and Krishan T,  The EM Algorithm and
Extension. {\it Wiley Sons, NY, 1997}.

\bibitem{erdos} Erdos P and Reyni A, The Art of Counting,
Edit. by J. Spencer (MIT Press, Cambridge MA , 1973).

\bibitem{haggai1} Kfir H and Kanter I (unpublished).















%



%
%
%
%
%
%
%
%























































%

%




\end{thebibliography}
\end{document}